\documentclass[conference]{IEEEtran}
\pdfoutput=1 

\usepackage{balance}

\usepackage{amssymb}
\usepackage{amsmath}
\usepackage{csquotes}
\usepackage{paralist}
\usepackage{url}
\usepackage[plain]{algorithm}
\usepackage{graphicx}
\usepackage{fixltx2e}
\usepackage{algpseudocode}
\usepackage{comment}
\usepackage{subcaption}
\usepackage{color}
\definecolor{light-gray}{gray}{0.35}

\newcommand{\tuple}[1]{\ensuremath{\left \langle #1 \right \rangle }}

\MakeRobust{\Call}

\newcommand{\Paragraph}[1]{\smallskip\noindent{\bf #1.}}

\algnewcommand\algorithmicswitch{\textbf{switch}}
\algnewcommand\algorithmiccase{\textbf{case}}
\algnewcommand\algorithmicassert{\texttt{assert}}
\algdef{SE}[SWITCH]{Switch}{EndSwitch}[1]{\algorithmicswitch\ #1\ \algorithmicdo}{\algorithmicend\ \algorithmicswitch}%
\algdef{SE}[CASE]{Case}{EndCase}[1]{\algorithmiccase\ #1}{\algorithmicend\ \algorithmiccase}%
\algtext*{EndSwitch}%
\algtext*{EndCase}%

\algrenewcommand\algorithmicindent{0.5em}%

\title{
Automatic Generation of Security Argument Graphs 
}

\author{
\IEEEauthorblockN{Nils Ole Tippenhauer}
    \IEEEauthorblockA{\\Singapore University\\ of Technology and Design\\
    nils\_tippenhauer@sutd.edu.sg}\\
\and
\IEEEauthorblockN{William G. Temple, An Hoa Vu,\\ Binbin Chen}
    \IEEEauthorblockA{Advanced Digital Sciences Center,\\ Singapore\\
    \{william.t, anhoa.vu, binbin.chen\}@adsc.com.sg}\\
\and
\IEEEauthorblockN{David M. Nicol, Zbigniew Kalbarczyk,\\ William H. Sanders} 
    \IEEEauthorblockA{University of Illinois \\at Urbana-Champaign, USA\\
    \{dmnicol, kalbarcz, whs\}@illinois.edu}
}

\begin{document}

\maketitle

\begin{abstract} 
  Graph-based assessment formalisms have proven to be useful in the
  safety, dependability, and security communities to help stakeholders
  manage risk and maintain appropriate documentation throughout the
  system lifecycle. In this paper, we propose a set of methods to automatically
  construct {\em security argument graphs}, a graphical formalism that integrates various security-related information to argue about the security level of a system. Our approach is to generate the graph in a progressive manner by exploiting logical relationships among pieces of
  diverse input information.  Using those emergent \emph{argument
    patterns} as a starting point, we define a set of \emph{extension
    templates} that can be applied iteratively to grow a security
  argument graph. Using a scenario from the electric power sector, we
  demonstrate the graph generation process and highlight its
  application for system security evaluation in our prototype software
  tool, CyberSAGE.

\end{abstract}



\section{Introduction}
\label{sec:intro}
Critical public infrastructure systems, such as those found in the
electric power and water sectors, must operate safely and reliably for
decades.  During their operating lifetimes, these systems are often
modified to face evolving operating conditions and requirements. 
For example, infrastructure systems have adopted greater communication and control 
capabilities in recent years. 
However, while these advanced features enable greater system
visibility and more efficient control strategies, they also open new
avenues for malicious attacks on the
system~\cite{amin13cyber1,anderson10,temple13delay}.

To understand evolving system requirements, operational contexts, and/or security threats, practitioners often employ graph-based reasoning techniques. Such approaches include safety cases~\cite{kelly98,denney12,denney13}, fault tree analysis~\cite{vesely81fault,pai02,xiang11}, and attack trees/graphs~\cite{schneier99,sheyner02,ou06,lemay11,kordy10adt}. Historically, development and maintenance of those graphical approaches required significant human effort.  
Recently, several efforts have begun in the safety and reliability communities to automate those processes~\cite{denney12,denney13,pai02,xiang11}. However, in the security domain, 
automation has been largely restricted to specific applications, such as construction of attack graphs~\cite{ou06}. The various challenges about security assessment were discussed in, e.g., \cite{nicol04, verendel09}.

Our approach for conducting holistic security assessment is to develop {\em security argument graphs}, a graphical formalism that integrates diverse inputs---including workflow information for processes executed in the system, physical network topology, and attacker models---to argue about the level of security for the target system. 
In our earlier work~\cite{Chen2013}, we presented an integrative security assessment framework that reasons about security by progressively combining heterogeneous types of information to construct such holistic security argument graphs. Section~\ref{sec:background} will recap the proposed framework and describe the unique structure of the generated security argument graphs.

Such holistic security argument graphs are beneficial in multiple ways: they make explicit the functional interdependecies of different pieces of security-related information;
also, the graph structure can be used to combine various numerical evidence to yield holistic quantitative security metrics.

In this paper, we provide a rigorous set of methods for constructing the holistic argument graphs introduced in~\cite{Chen2013}. We leverage recurring \emph{argument patterns} that emerge from the need to integrate heterogeneous information about a system 
and possible attacks, 
and instantiate them as \emph{extension templates} that can be iteratively and automatically applied to grow our argument graphs. Using our Cyber Security Argument Graph Evaluation (CyberSAGE) tool~\cite{cybersage}, which is currently under development, we demonstrate the automated graph generation for an example electric power grid system~\cite{etsi-architecture}.

The remainder of this paper is structured as follows: Section~\ref{sec:background} reviews our assessment framework and the structure of the resulted security argument graphs. In Section~\ref{sec:patterns}, we
present the argument patterns that can be used to generate such
argument graphs. In Section~\ref{sec:construction}, we formalize the
argument patterns as extension templates. In
Section~\ref{sec:application}, we show how to use these extension templates to
automate the argument graph construction process. A practical example of the graph
construction is presented in Section~\ref{sec:usecase}, with a use
case from the electric power sector.  We discuss related work in
Section~\ref{sec:related}, and conclude in
Section~\ref{sec:conclusion}.

\section{Security Argument Graphs}
\label{sec:background}

Many safety~\cite{kelly98}, reliability~\cite{vesely81fault}, and
security~\cite{kordy2013survey} assessment methodologies rely on graphical
structures to organize and present information. In a security context,
such ``argument graphs"~\cite{nicol04} could help provide a precise
underpinning for threat modeling and quantitative evaluation of
system-level metrics. A particular challenge in constructing security
argument graphs is that of dealing with the heterogeneous set of
information that needs to be incorporated, which may include security
requirements, business processes, system architecture, physical device
specifications, known vulnerabilities, and attacker models.

\begin{figure}[tb]
\centering 
  \includegraphics[width=0.9\linewidth]{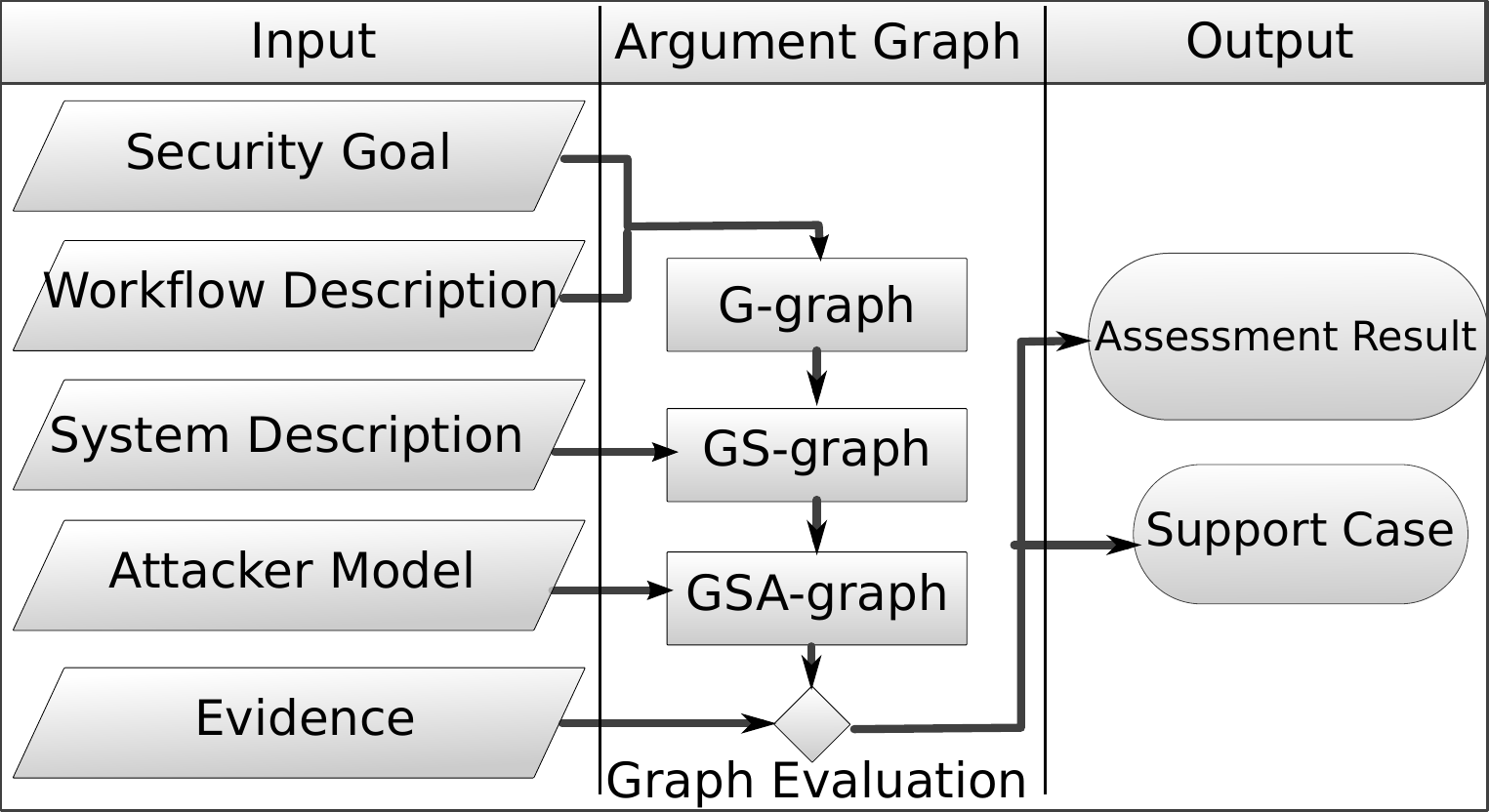}
  \caption{The security assessment framework proposed in~\cite{Chen2013}.}
  \label{fig:framework}
\end{figure} 

In~\cite{Chen2013}, we presented a high-level workflow-oriented security assessment
framework, which organizes the diverse set of information inputs described above. The
overall assessment process, which is illustrated in
Fig.~\ref{fig:framework}, relies on a unique argument graph structure
that progressively incorporates Goal, System, and Attacker (GSA)
information.

As a brief overview, the process starts with a precisely defined
security goal, which may relate to properties such as Confidentiality,
Integrity, or Availability. The analyst then identifies system
processes that are relevant to the security goal, and represents them
as workflow diagrams. Those workflow diagrams provide a sequence of
actions and their respective actors. Those two inputs are combined to
form a simple argument graph, called a Goal (G) graph, that captures
information about actors and interactions that may affect the security
goal. When detailed system information (e.g., actor to device mapping,
network topology, or device configuration) is available, we use the
G-graph to integrate that information and
generate a more detailed security argument graph: the Goal, System
(GS) graph.  Finally, we incorporate information about possible
attacker actions and capabilities into the GS-graph to generate the
Goal, System, Attacker (GSA) graph.

\begin{figure}[tb]
\centering 
  \includegraphics[width=\linewidth]{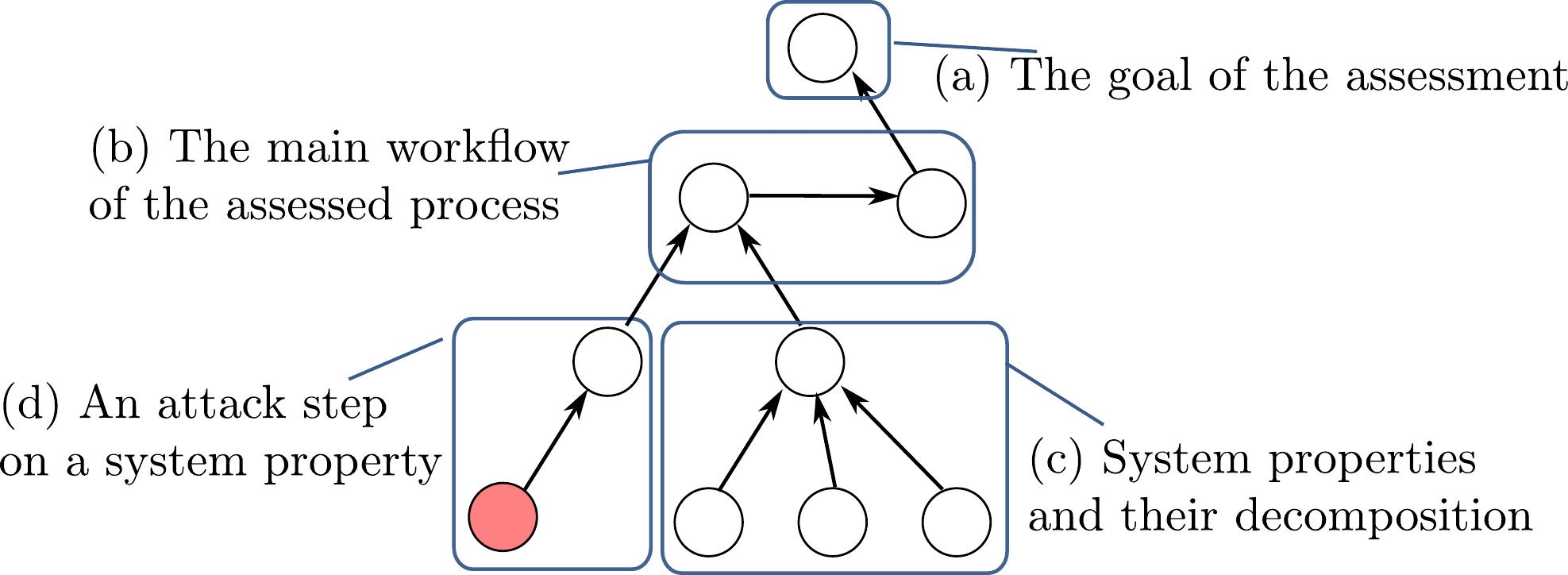}
  \caption{Basic components of the argument graph: (a) the workflow steps of the assessed process, (b) a system and device property and its decomposition, and (c) a potential attack step and its immediate connection to system property.}
  \label{fig:overviewStructure}
\end{figure}

Our GSA graph structure, which is represented in
Fig.~\ref{fig:overviewStructure}, is system-focused (like fault
trees~\cite{vesely81fault}), but allows for the modeling of attacks
(like attack graphs~\cite{sheyner02}). The graph contains
vertices of different types, such as system (attacker) actions and
system (attacker) properties.  This graph structure has no explicit
vertices to denote aggregation semantics (e.g., OR relations or AND
relations), as each vertex contains information that defines the
aggregation of its incoming neighbors. Thus, the graph has only a
single type of dependency relationship among the vertices.

\section{Argument Patterns in Security Assessment}
\label{sec:patterns}
While applying our structured approach to generate security argument
graphs (e.g., in the context of smart grid infrastructure), we
observed a series of recurring {\em argument patterns}.  These
patterns capture direct logical relationships among different pieces
of information and may be used to develop reusable {\em extension
  templates} that help automate the graph generation process.  In this
section, we give an overview of several useful argument patterns we
identified in our work on security assessments of processes in the
smart grid domain. We shall see that the patterns to be presented are
fairly generic, so they are applicable to other domains as well. In
addition, other domains might also contribute patterns that we have
not yet encountered. We formalize those patterns as extension
templates in Section~\ref{sec:rewriting}.

In general, two classes of patterns occur in our arguments:
\emph{intra-type} patterns, and \emph{inter-type} patterns. As the
name suggests, intra-type patterns introduce and connect vertices of
the same type, and inter-type patterns introduce and connect vertices
of different types.  The following describes the two pattern classes in more
detail and provide five example patterns we identified. We number the
patterns according to the order in which they typically appear in
our argument graphs in the generation process.

\Paragraph{Inter-Type patterns}
Inter-type operations connect vertices of different types. For
example, at a high level, a security goal is directly defined in the
context of one or more workflows to which the goal is related. We
characterize such a relationship by the following argument pattern $P_1$:
\emph{security goals or requirements directly depend on processes that
  occur in the system.} As another example, consider the workflow
steps in our argument graph. Each workflow step is performed by one
or multiple actors with certain properties. This association of
workflow steps with actors is our argument pattern $P_3$: \emph{to successfully complete actions in the
  workflow, their respective actors need to be available.}
In addition, these abstract actors have to be mapped to concrete
devices in the system to allow a more detailed decomposition. That is
our argument pattern $P_4$: \emph{devices in the system adopt one or more
  workflow actor roles to provide functionality.}

\Paragraph{Intra-Type patterns}
Workflows in the system have a number of actions that have to be
executed in sequence. Our graph generation starts with the final step
of a workflow, and then adds its prerequisite step (i.e., the
generation works backwards in time). That is one of our argument
patterns, pattern $P_2$: \emph{the successful completion of workflow actions may depend on 
the other preceding or concurrent actions.} For example, assume
that the final step in a workflow depends on the receiving of a
message. The intra-type pattern can be used to identify the workflow
step that is related to the transmission of that message. Another
important intra-type pattern relates to specific devices (within a
system) that may be involved in a workflow. The refinement of device
properties is pattern $P_5$: \emph{device properties depend on
  sub-properties and their composition semantics.} For example, the
availability of a device depends on the availability of its software
and hardware, so this property can be decomposed.

In general, our argument patterns capture individual direct logical
relationships that make up the entire argument. Such a focus on simple
patterns of abstract relationships allows us to apply the patterns
with only local knowledge of the argument graph. As each argument
pattern captures a generic relationship, we can construct our argument
graph using a small number of patterns. For an analogy, argument
patterns function like axioms that could be used to build up a
mathematical proof. Like axioms, argument patterns are
derived from logical relationships, best practices, common scenarios, and
expert knowledge.  Thus, different patterns or different ways of
instantiating the same patterns are typically required to tackle
security problems of different nature, (e.g., availability
vs. confidentiality). The patterns presented in this paper are
used to support an availability assessment use case (see
Section~\ref{sec:usecase}). 

In the next section, we show how to formalize those argument patterns by constructing \emph{extension templates}. The resulting set of extension templates allow us to automatically generate an argument graph based on several classes of inputs which drive the security assessment.

\section{Graph Generation Based on Argument Patterns}
\label{sec:construction}

So far, we have introduced the concept of argument patterns that
emerge during the construction of argument graphs; we also provided several informal
examples to illustrate our intuition.
We now present a formalism that rigorously defines these patterns and the manner in which they can be applied. 
In the following, we start with a formal definition of the security argument graphs. We then formalize how to progressively generate such graphs through the application of \emph{local extensions}. We define how a \emph{local extension} is generated through the instantiation of an \emph{extension template}, which formalizes a corresponding argument pattern. With all these building blocks in place, we then present our overall process for generating the graph.


\subsection{Graph Structure and Local Extensions}
\label{sec:rewriting}

We first define the structure of our graph and how a graph can be generated by the application of local extensions.

\Paragraph{Graph Definition}
For the following discussion, a graph $\omega_i$ is defined to be a
triple $$\omega_i:=\tuple{V_i, E_i, l_i},$$ where $V_i$ is a finite
set of vertices, $E_i$ is a finite set of directed edges, and $l_i$ is
the labeling function. Each vertex is itself a static tuple that
contains the type of the vertex, and some of type-specific additional
data. The type and data are set when the vertex is created, and cannot
be changed afterward. Two vertices are considered identical whenever
all their static data are identical. An edge $e=\tuple{v_s, v_t}$ is
represented by a tuple that contains references to its source vertex
$v_s$ and target vertex $v_t$. The labeling function, $l_i(v)$,
returns the mutable attribute(s) of a vertex $v \in V_i$. For example,
a variable attribute could be the probability that this vertex's
property is true (to be determined in the graph evaluation
later). Note that $l_i(v)$ does not need to be a single numerical
value: it can encapsulate a list of different types of information,
such as an expression relating the attributes of its incoming
neighbors to its own
attribute. 

\Paragraph{Local extension}
We progressively generate the security argument graph through the application of \emph{local extensions}. A local extension $r$ is defined as a
tuple of its matching vertex $v_r$ and the resulting star graph
$\omega_r$:
$$r:=\tuple{v_r,\omega_r}$$

The resulting star graph $\omega_r$ contains $v$ and at least one additional vertex. Each of these additional vertices has one outgoing edge towards $v$. Other than those edges, there is no other edge in $\omega_r$.

\begin{figure}[tb]
\centering
  \includegraphics[width=0.8\linewidth]{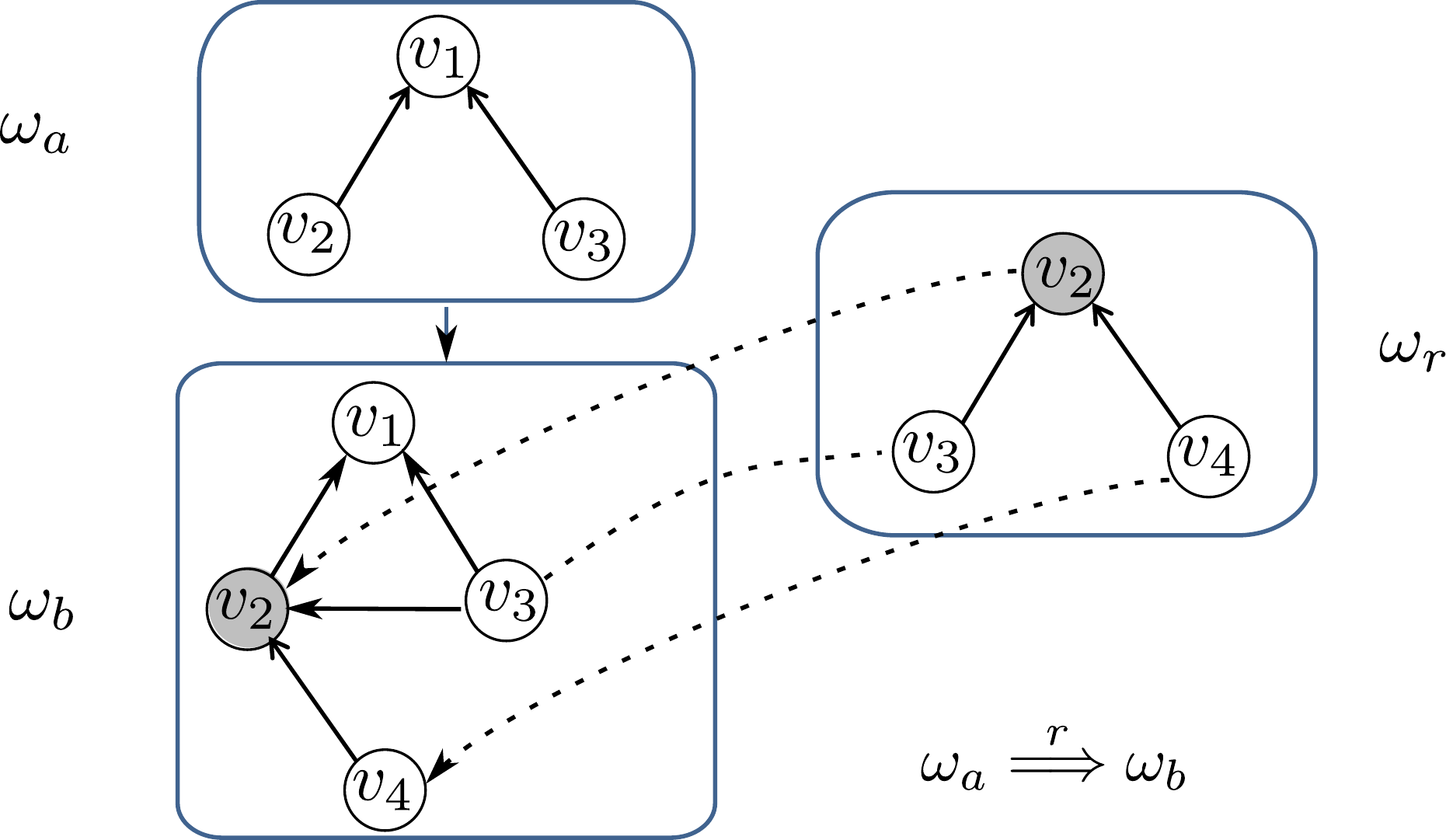}
  \caption{Local extensions: Transformation of graph $\omega_a$ using local extension $r$ into graph $\omega_b$ (i.e. $\omega_a \stackrel{r}{\Longrightarrow}\omega_b$). The local extension $r$ matches $v_2$, and inserts $\omega_r$ in its place. Slashed arrows denote logical connections between graphs. For illustration purposes, the variable attribute of a vertex is its color.}
  \label{fig:apply-decomposion-rule}
\end{figure}

We use $\omega_a\stackrel{r}{\Longrightarrow }\omega_b$ to denote the application of a local extension $r$ to a graph $\omega_a$, which generates a new graph $\omega_b$ (see Fig.~\ref{fig:apply-decomposion-rule}). 
Here, we assume that $r$ is applicable, i.e., the matched vertex $v_r$ is indeed a vertex of $\omega_a$.  Given that, the local extension is applied as follows:
$$\omega_a\stackrel{r}{\Longrightarrow }\omega_b = \tuple{V_a \cup V_r,E_a \cup E_r,l_b}$$
where
$$l_b(v) =  \begin{cases}
l_r(v) & \text{if } v \in (V_r \setminus V_a) \cup \{v_r\} \\
l_a(v) & \text{otherwise.}
\end{cases}$$

The additional vertices from the star graph $\omega_r$ and the associated edges are added to the original graph. Note that the additional vertices in $\omega_r$ may or may not be present in the original graph $\omega_a$. For each vertex that is already present in the original graph $\omega_a$, except for $v$, the old labeling function is preserved; otherwise, the labeling function is taken from the star graph.

\subsection{Extension Templates and Graph Generation}
Having introduced local extensions, we now consider their
generation. We introduce the notion of an
\emph{extension template,} $$\gamma:=\tuple{m_\gamma, f_\gamma},$$ where
$m_\gamma$ is a \emph{matching function} and $f_\gamma$ is an
\emph{extension generation function}. Specifically, $m_\gamma(v,
\Sigma)$ shows $\gamma$'s matching score for vertex $v$, where $\Sigma$
is the \emph{environment}: a placeholder for additional information. Recall that
the attribute of a vertex $l(v)$ is a tuple that contains its type and
type-specific additional information, both of which can be used to
determine the numerical value of $m_\gamma(v, \Sigma)$. A value of
$m_\gamma(v, \Sigma)=0$ means that extension template $\gamma$ is not applicable to
the generation of an extension for $v$. If multiple extension templates are applicable
(i.e., several extension templates can be applied to the same vertex), then
$m_\gamma$ could be implemented as chosen from a range, to indicate
precedence. Otherwise, $m_\gamma$ can be implemented as a Boolean
function indicating applicability of $\gamma$.

If an extension template $\gamma$ is applicable to a vertex $v$, the local extension
generation function $f_\gamma(v,\Sigma)$ uses the information from
vertex $v$ and environment $\Sigma$ to generate a local extension
that can be used to expand $v$. In that case, $\Sigma$ is a
placeholder for various pieces of information that are relevant to the
transformation of the graph, such as the workflow information,
the actor-to-component mapping, or the network topology graphs. We use $\Gamma$ to 
denote the set of extension templates.

\Paragraph{Graph Generation Algorithm}
We define \emph{graph generation} as the repeated application of a
set of extension templates to a graph. More formally, it is the application of
$\Gamma$ to a graph $\omega_a$, using the environment $\Sigma$. The
underlying algorithm is summarized as pseudocode in
Algorithm~\ref{alg:graphA} below.

\begin{algorithm}[H]
\centering
  \includegraphics[width=0.8\linewidth]{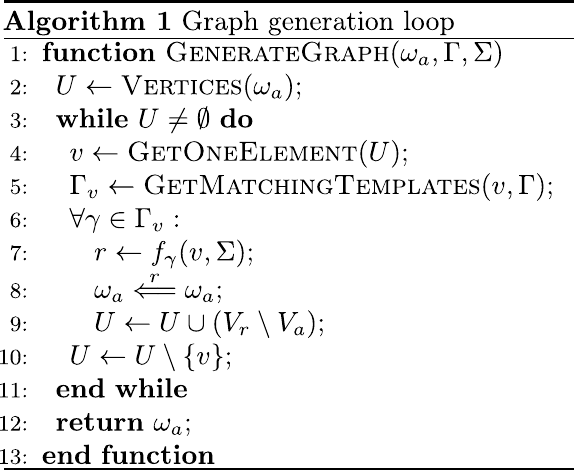}
\caption{Graph generation loop}\label{alg:graphA}
\end{algorithm}

In the pseudocode, $\Call{Vertices}{\omega_a}$ returns all vertices of
the graph, while $\Call{GetMatchingTemplates}{v, \Gamma}$ returns a
set $\Gamma_v$ of all templates applicable to $v$, and
$\Call{GetOneElement}{U}$ simply picks an arbitrary element of the set
$U$. Note that when no template is applicable, $\Gamma_v = \emptyset$,
whereas when several templates are applicable, the one with the highest matching
score $m_\gamma$ is chosen.  In Table~\ref{tab:templates}, we list some
of those extension templates and describe their functions. The
template numbering corresponds to the pattern numbering in
Section~\ref{sec:patterns}; the set has been expanded to include two
additional templates, which relate to attacker modeling.  Here, we
concentrate on extension templates related to availability of a
process, matching our case study in Section~\ref{sec:usecase}. We are
currently working on additional extension templates to model the
transmission path of messages, human-machine interactions, and more
complex workflow
mechanisms. 

\begin{table}[t]
\begin{center}
\begin{tabular}{ l| l }
  ID & Comment  \\\hline
  $T_1$ &  Template to connect goal node to assessed workflow(s)  \\
  $T_2$ &  Template to look up required previous steps in a workflow \\
  $T_3$ &  Template to create requirements for the actor of a workflow step \\
  $T_4$ &  Template to create device requirements for an actor \\
  $T_5$ &  Template to decompose requirements for devices \\
  $T_6$ &  Template to identify potential attacks on leaf properties  \\
  $T_7$ &  Template to create requirements for an attack step  \\ \hline
\end{tabular}
\caption{A sample set of extension templates.}
\label{tab:templates}
\end{center}
\end{table}

\section{Using Extension Templates to Generate Argument Graphs}
\label{sec:application}

Our security assessment process~\cite{Chen2013}, shown in Fig.~\ref{fig:framework}, contains three types of security argument graph: the G-graph, GS-graph, and GSA-graph.  All of those graphs are generated through application of Algorithm~\ref{alg:graphA}; the differences arise from the specific inputs provided and the extension templates used for graph generation. The meta-process for constructing the argument graphs is:
$$
\begin{array}{r l}
\omega_g &\gets \Call{GenerateGraph}{\omega_0, \Gamma_g, \Sigma_g}\\
\omega_{gs} &\gets \Call{GenerateGraph}{\omega_g, \Gamma_s, \Sigma_s}\\
\omega_{gsa} &\gets \Call{GenerateGraph}{\omega_{gs}, \Gamma_a, \Sigma_a}
\end{array}
$$
In the following three subsections, we describe individual steps,
their inputs, and the associated templates in greater detail. Finally,
in Section~\ref{sec:T5example}, we focus on a single template to
illustrate the level of specification required, and the importance of
templates in automating the graph generation process.

\subsection{Graph Generation Using Workflow Input}
To generate the first stage of our argument graph, the G-graph,
Algorithm~\ref{alg:graphA} is called as follows:
$\Call{GenerateGraph}{\omega_0, \Gamma_g, \Sigma_g}$. That will
generate the G-graph, which includes the workflow input in $\Sigma_g$,
and is based on an initial base-graph $\omega_0$ and a set of
extension templates $\Gamma_g$. The initial graph will contain only a vertex representing the goal of the assessment. $\Gamma_g$ contains $T_1,T_2,$ and
$T_3$, the formal extension templates for patterns $P_1,P_2,$ and $P_3$ as defined in
Section~\ref{sec:patterns}.

\Paragraph{Workflow information in $\Sigma_g$}
The workflow input describes \emph{how the system provides a
  functionality}, and identifies necessary actors as well as the
information they exchange. Our internal structure for this input is a
workflow graph: $\omega_w=\tuple{V_w,E_w,l_w}$. For $v \in V_w$,
$v=\tuple{type, attributesTuple}$. Workflow vertices have only one
type of vertex: $\Call{Type}{v} \in \{\text{WorkflowStep}\}$.  A
vertex's attributesTuple field depends on its type.  For simplicity,
in the following we consider only sequential workflows, with workflow
step vertices. The attribute tuple of each workflow step vertex has
two static attributes: {\tt actor} and {\tt action}. Thus, the attributeTuple is
$\tuple{actor, action}$. The labeling function $l_w$ is used to store
additional workflow information, e.g., on branching, merging, and
conditions in the workflow.

\Paragraph{Argument graph vertex types and templates $\Gamma_g$} 
We construct our argument graph iteratively, starting with $\omega_0$,
which is a directed graph with the assessment goal as its only
vertex. The assessment goal determines which properties of the system
we are interested in, for example, the availability of the components
involved in a workflow. The goal vertex is directly created from user
input, without any need for an extension template. This goal vertex is
then connected to the final step of the assessed workflow, using
extension template $T_1$ (which implements $P_1$: see
Section~\ref{sec:patterns}).  Then, that final workflow step vertex is
connected to the required previous steps in the workflow using $T_2$
(implementing $P_2$), and those steps are expanded as well. All
vertices generated by $T_1$ and $T_2$ will be of type
``ActionAvailability.'' In addition, required properties for each
actor, such as actor and communication link availability, are added
using extension template $T_3$ (implementing $P_3$).  That will add
vertices of type ``ActorAvailability'' and ``MessageAvailability.''  A
vertex with {\tt type} ``ActionAvailability'' has static attributes
{\tt action} and {\tt actor}. For example, in the context of the smart grid, the {\tt action} can be
``MeterReading,'' and the {\tt actor} can be ``Utility''. A vertex with
{\tt type} ``ActorAvailability'', has static attribute {\tt
  actor}. For example, the {\tt actor} can be ``Utility.''  We call
the resulting graph at this stage the \emph{G-graph}, as it models the immediate
requirements related to the goal of the assessment.

\subsection{Graph Generation Using System Input}
\label{sec:systeminput}
As the next step, the argument graph $\omega_G$ is expanded using
Algorithm~\ref{alg:graphA}, this time with different operands:
$\Call{GenerateGraph}{\omega_g, \Gamma_s, \Sigma_s}$. Here, the
operands are the system information $\Sigma_s$ and a set of extension templates
$\Gamma_s$.  Here, $\Gamma_g$ contains $T_4$ and $T_5$, the formal
extension templates for patterns $P_4$ and $P_5$.

\begin{figure}[tb]
\centering 
  \includegraphics[width=\linewidth]{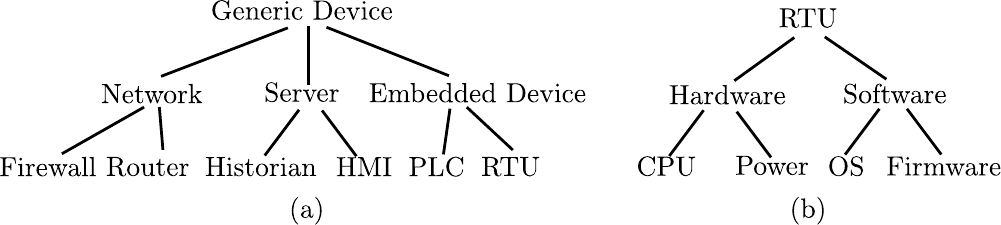}
  \caption{Examples of a device type hierarchy (a) and composition tree (b). The device types are based on the power grid substation example, with an HMI (Human-Machine Interface), PLC (Programmable Logic Controller), and RTU (Remote Terminal Unit). } 
  \label{fig:combinedTrees}
\end{figure}

\Paragraph{System information inputs} The system description contains multiple types of inputs, 
including the following. 
\begin{compactitem}
\item An actor-to-component mapping: This is similar to the deployment diagram in UML. It maps the actor from the workflow to a {\tt componentType}. 
\item A network topology graph $\omega_n=\tuple{V_n,E_n, l_n}$, where each vertex $v \in V_n$ is a physical device in the system, and each edge $e \in E_n$ is a link (which can be a single-hop physical communication link, a network path, or physical). The attribute function $l_n(v)$ describes various attributes of a device $v$, including its type, physical location, and access privileges, among others. For a link $e$, $l_n(e)$ describes its attributes, like type, capacity, and delay. 
\item A {\tt componentType} hierarchy diagram as depicted in  Fig.~\ref{fig:combinedTrees}a. 
\item The device composition information as depicted in Fig.~\ref{fig:combinedTrees}b.
\end{compactitem}

\Paragraph{Vertex types and templates for GS-graph generation} Based
on those inputs, two extension templates can be applied to the argument graph:
$T_4$ and $T_5$. $T_4$ makes it possible to connect the abstract actor in the
workflow with the concrete device that executes this action,
thus introducing ``ComponentAvailability'' vertices to the graph. A
vertex of that type has 2 static attributes: {\tt component} and {\tt
  componentType}. The {\tt componentType} is taken from a hierarchy of
device types with increasing specificity towards the graph's leaves. The {\tt
  component} is taken from a tree that describes the subcomponents
for each {\tt componentType}. If no subcomponent tree is available for
a given {\tt componentType}, the subcomponent tree of a parent {\tt
  componentType} will be used instead. We describe $T_5$ in more
detail in Section~\ref{sec:T5example}.

\subsection{Graph Generation Using Attacker Input}
As the last step, the argument graph $\omega_{gs}$ is expanded using
Algorithm~\ref{alg:graphA}, this time with different operands:
$\Call{GenerateGraph}{\omega_{gs}, \Gamma_a, \Sigma_a}$. Here, the
operands are the system information $\Sigma_a$ and a set of extension templates
$\Gamma_a$. The two patterns for the extension templates $T_6$ and $T_7$ in
$\Gamma_a$ were not introduced earlier. They are similar to $P_2$ and $P_3$,
but relating to the attacker (instead of the system).

\Paragraph{Attacker information input} The input placeholder
$\Sigma_a$ here contains the attacker model, which contains a set of
attacker properties and a set of attack sequences $\Omega_r$ (a set
of star graphs). Each star graph $\omega_r$ contains a potential
attack step and its immediate prerequisites. The vertices in
$\omega_r$ are of {\tt type} ``AttackStep'' and
``AttackerProperty.'' Attacker property information relates to methods,
knowledge, and physical access of the attacker, e.g., access to a
company's compound or server room.

\Paragraph{Vertex types and templates for GSA-graph generation} In the
GSA-graph, new vertices of {\tt type} ``AttackStep'' and
``AttackerProperty'' are introduced (matching the nodes from the
attacker model). Under this model, only single step attacks and their corresponding requirements can be modeled, however we are developing additional vertex types to incorporate multi-step attacks.

\subsection{Automatic Generation using Templates}
\label{sec:T5example}

\begin{figure}[tb]
\begin{subfigure}{\linewidth}
\centering
\includegraphics[width=0.8\linewidth]{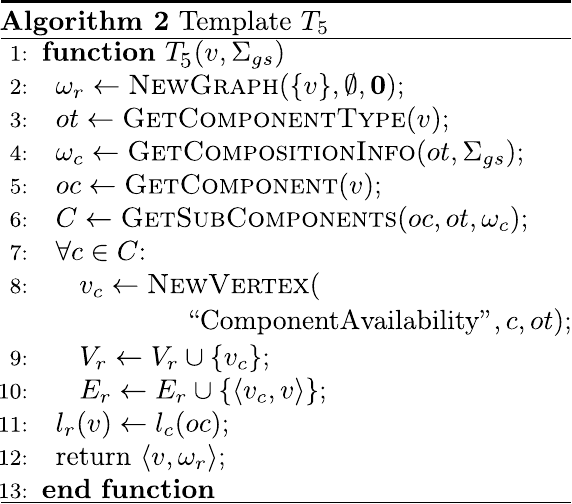}
\caption{}
\label{fig:t5}
\end{subfigure}

\begin{subfigure}{\linewidth} 
\begin{center}
  \includegraphics[width=0.7\linewidth]{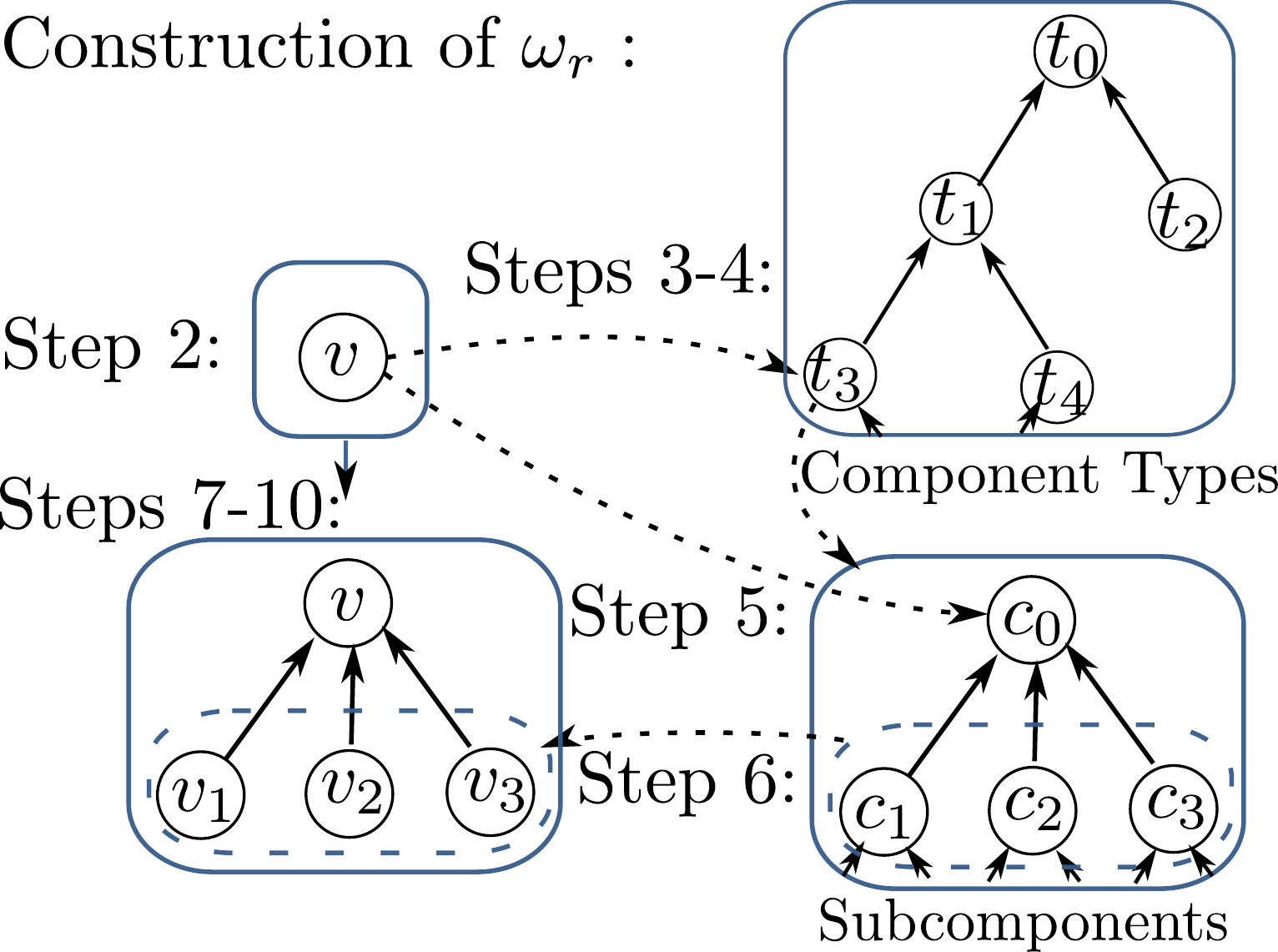}
\caption{}
\label{fig:exampleComponentAvailability}
\end{center}
\end{subfigure}
  \caption{Details on extension template $T_5$: (a) Pseudocode of the extension function \Call{$T_5$}{$v,\Sigma_a$}. (b) Visualization of the process.}
  \label{fig:t4example}
\end{figure}

In the previous sections, we have introduced several extension
templates that we distilled from common argument patterns that we
observed. As part of our effort toward automating the security
assessment of complex systems, we are compiling an extension template
library. All of the extension templates in Table~\ref{tab:templates}
are defined in pseudocode. In this paper, we present only the core set
of templates we are currently working on: additional extension
templates are omitted from this paper because of space limitations.
We use the extension template $T_5$ ({\em device decomposing requirements}) to illustrate the underlying
extension process.


Extension template $T_5$ relies on component type and device
composition hierarchies, which are specified as an input to the
security assessment (part of
$\Sigma_{gs}$). Fig.~\ref{fig:combinedTrees} presents examples of the
composition hierarchies, which were discussed in
Section~\ref{sec:systeminput}.  Template $T_5$ uses the supporting
hierarchies to expand a single graph vertex
$v$. 
We show the extension generation function in Fig.~\ref{fig:t5} and
represent the process graphically in
Fig.~\ref{fig:exampleComponentAvailability}. In particular, the local
extension for a component property is created by finding the best
matching {\tt componentType} from the {\tt componentType} tree. That best
{\tt componentType} is then used to find the matching property
composition tree, and to look up the next decomposition for the current
property. All potential decomposition nodes are added to a star graph
$\omega_r$, and returned together with the node $v$ as a local
extension.

That precise description of the extension template application process and required input information allows the extension templates to be readily implemented in a software tool. In the next section, we present an example security assessment, and show that the process can be automated using a supporting software tool that is currently under development.

\section{Application: An Electric Power Grid Use Case}
\label{sec:usecase}
In this section, we apply the algorithms and templates presented earlier to an illustrative use case from the electric power sector. We start by manually deriving a security argument graph to explain important details. We then discuss graph generation in CyberSAGE~\cite{cybersage}, a security assessment tool that is currently under development. Using CyberSAGE, we can generate the argument graph automatically, based on a library of extension templates and a set of inputs. 

\subsection{Assessment Input}

%

\begin{figure*}[!h]
\centering 
\begin{subfigure}{0.24\textwidth}
\centering
\includegraphics[width=\linewidth]{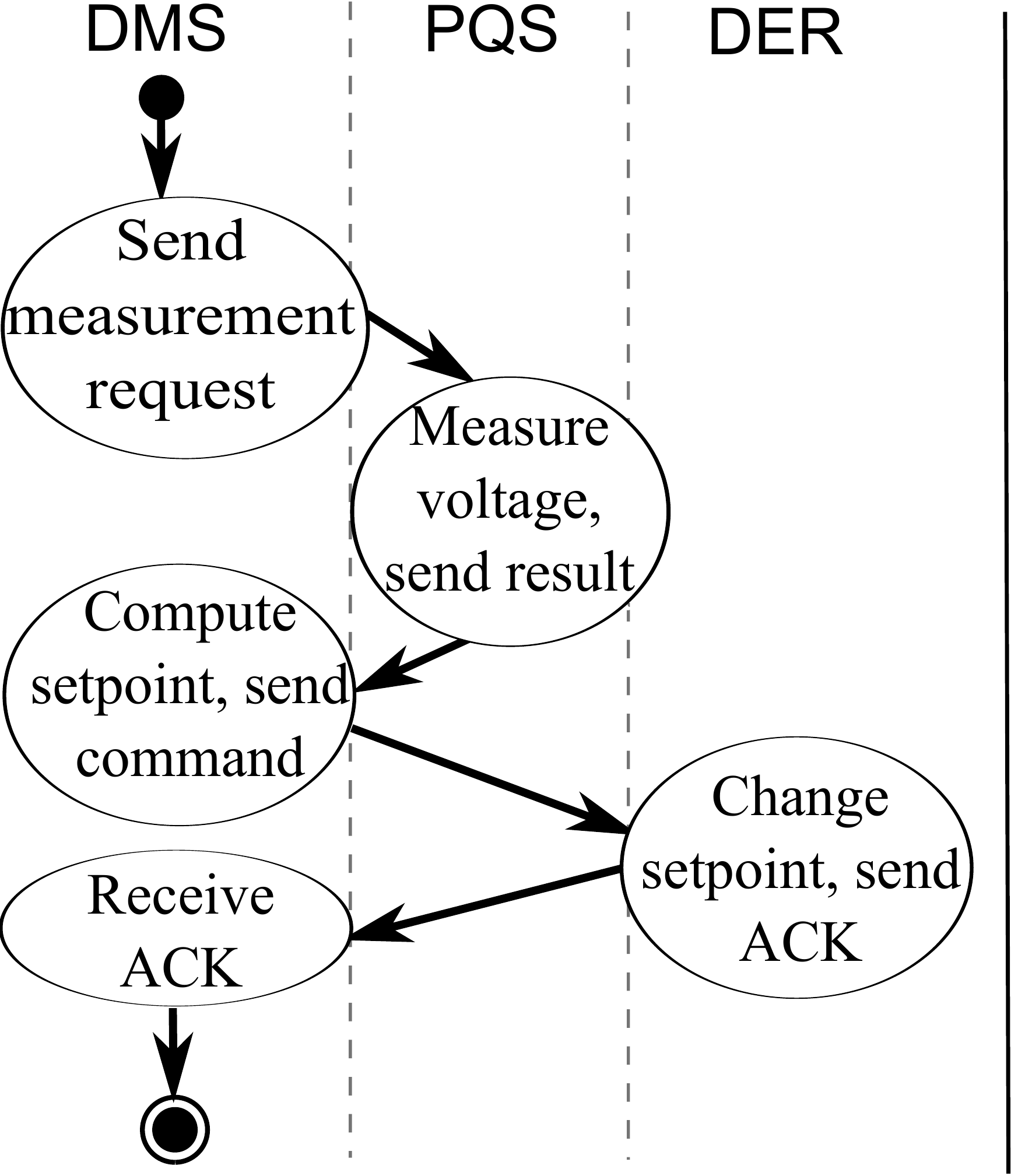}
\caption{}
\label{fig:workflow}
\end{subfigure}
\begin{subfigure}{0.74\textwidth}
\centering
\includegraphics[width=0.9\linewidth]{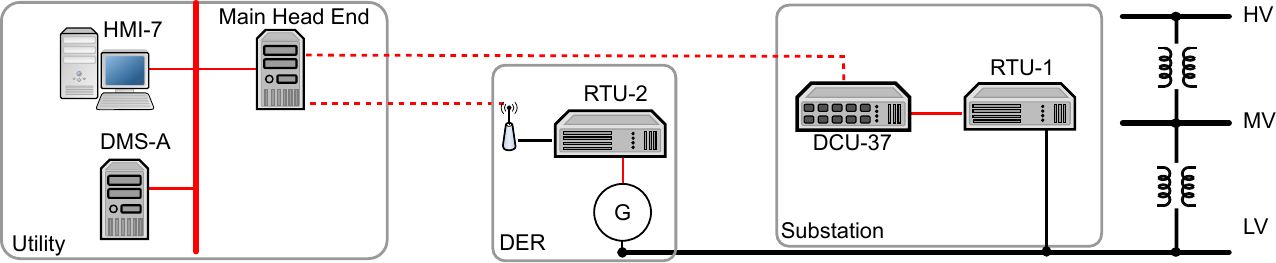}
\caption{}
\label{fig:system}
\end{subfigure}
\caption{Smart grid substation use case: (a) Voltage measurement and control workflow; (b) System topology.}
\label{fig:caseinputs}
\end{figure*}

We consider an example power system use case adapted from~\cite{etsi-architecture}. This scenario is a typical supervisory control and data acquisition (SCADA) operation, connecting a utility company's network with intelligent field devices that manipulate physical power grid parameters.
In this example, a central distribution management system (DMS) monitors the voltage at a specific point (i.e., bus) in the low voltage (LV) power distribution network as reported by the power quality sensor (PQS) there, and uses that information to trigger a control action in another device located in the network. The control device in this example is a distributed energy resource (DER), e.g., an electric vehicle or small wind turbine. 
An example workflow for the reactive power control process is shown in Fig.~\ref{fig:workflow}. 
Our security goal in this example is to ensure the availability of the measurement and control process. 
A simplified system diagram is shown in Fig.~\ref{fig:system}, which includes both communication links (red) and physical power network connectivity (black). Wide-area communication between different physical locations is indicated by dashed edges.
In the system shown in Fig.~\ref{fig:system}, RTU-1 is the device measuring the bus voltage, while RTU-2 is responsible for controlling the DER's reactive power output. Both devices are of \texttt{componentType} RTU. A server device (\texttt{componentType} ``server'') in the corporate network, DMS-A, is responsible for distribution management functionality.

The attacker model in this example includes attacks (i.e., ``AttackStep" vertices) targeting:
\begin{itemize}
\item Device power supply (Physical Tampering)
\item Device operating system (Exploit Vulnerability)
\item Device network connectivity (Denial of Service)
\end{itemize}
The ``AttackerProperty" information that enables these attack steps is specific to the class of device being targeted.

\subsection{Security Argument Graphs}
\begin{figure*}[!t]
\centering 
\includegraphics[width=0.8\linewidth]{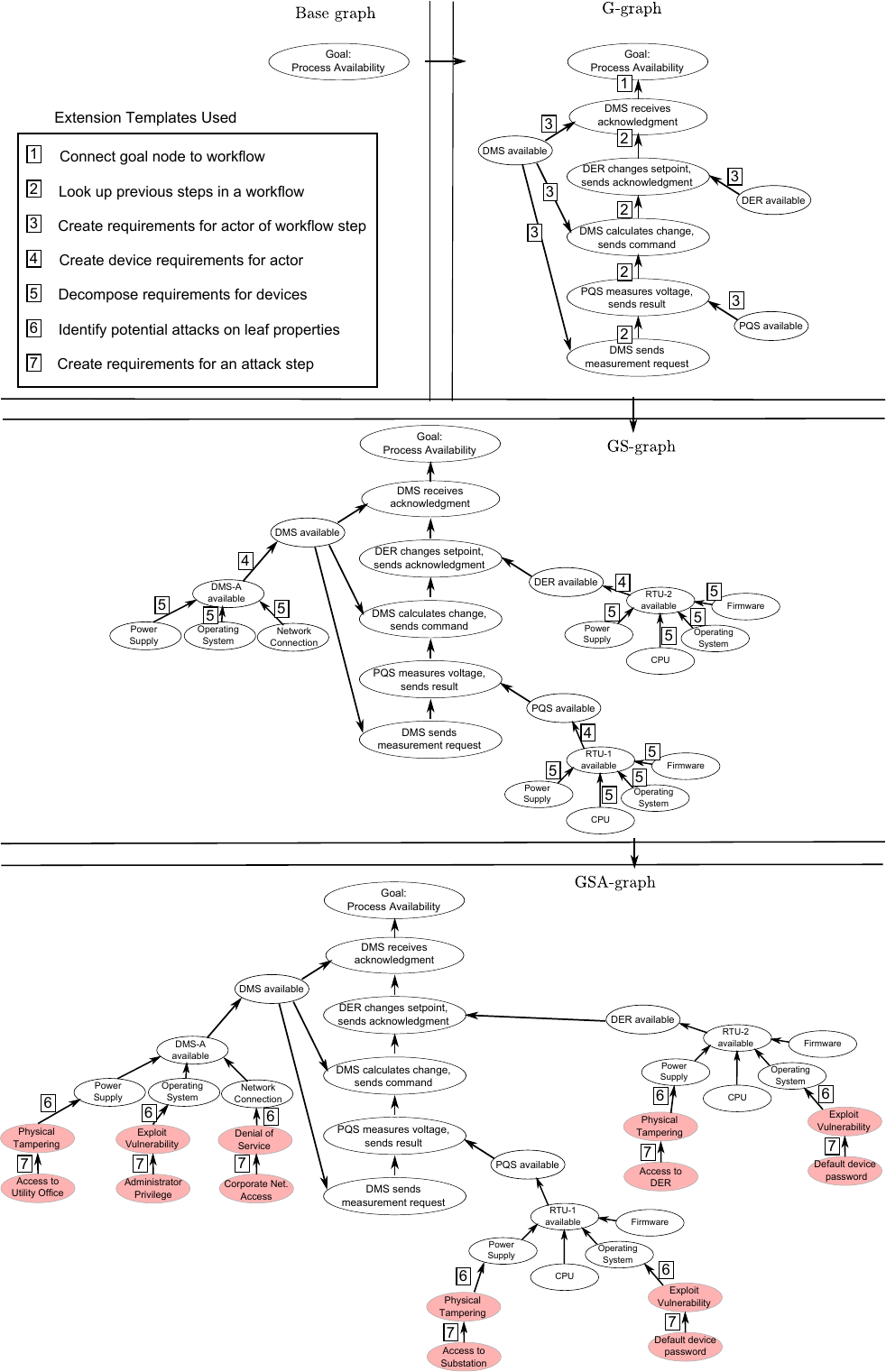}
\caption{Manually derived security argument graphs for the distribution automation use case. Each edge is annotated to show the extension template applied during graph generation.}
\label{fig:exampleGSA}
\end{figure*}

Knowledge of the assessment goal, the workflow (Fig.~\ref{fig:workflow}), and the system (Fig.~\ref{fig:system}) inputs, the actor \texttt{componentType} mapping, and an attacker model allows us to apply the extension templates defined in Table~\ref{tab:templates} to construct a security argument graph according to our framework~\cite{Chen2013}. 
The graph generation process is depicted visually in Fig.~\ref{fig:exampleGSA}. First, the base graph (representing the goal) is extended using workflow input. Extension templates $T_1$ to $T_3$ are used during this stage to identify dependencies between actions and actors. Next, templates $T_4$ and $T_5$ are used to enrich the argument graph with system-specific information (GS-graph). The fully developed GSA-graph, formed by applying $T_6$ and $T_7$, is shown at the bottom of Fig.~\ref{fig:exampleGSA}.

This complete security argument graph (GSA-graph) organizes security-related information that originated from disparate sources and formats. The human-readable structure is intended to help system designers and other stakeholders understand dependencies and possible security threats in a complex system. In particular, it is clear from Fig.~\ref{fig:exampleGSA} that the distribution management system (DMS) is critical to the voltage control process---it is involved in $3$ separate actions, occuring at different times. However, since the DMS is located in the utility's back-office network it may pose less of a security risk than the DER, which is a field device. While this analysis is qualitative, the GSA-graph can be used for quantitative security assessment as well. In~\cite{Chen2013} we outline an approach for quantitative evaluation over the argument graph. That quantitative evaluation has been refined and is implemented in our software tool, which is discussed below.


\subsection{Automation in the CyberSAGE Tool}

\begin{figure*}[!h]
\centering 
\begin{subfigure}{0.6\textwidth}
\centering
\includegraphics[width=\linewidth]{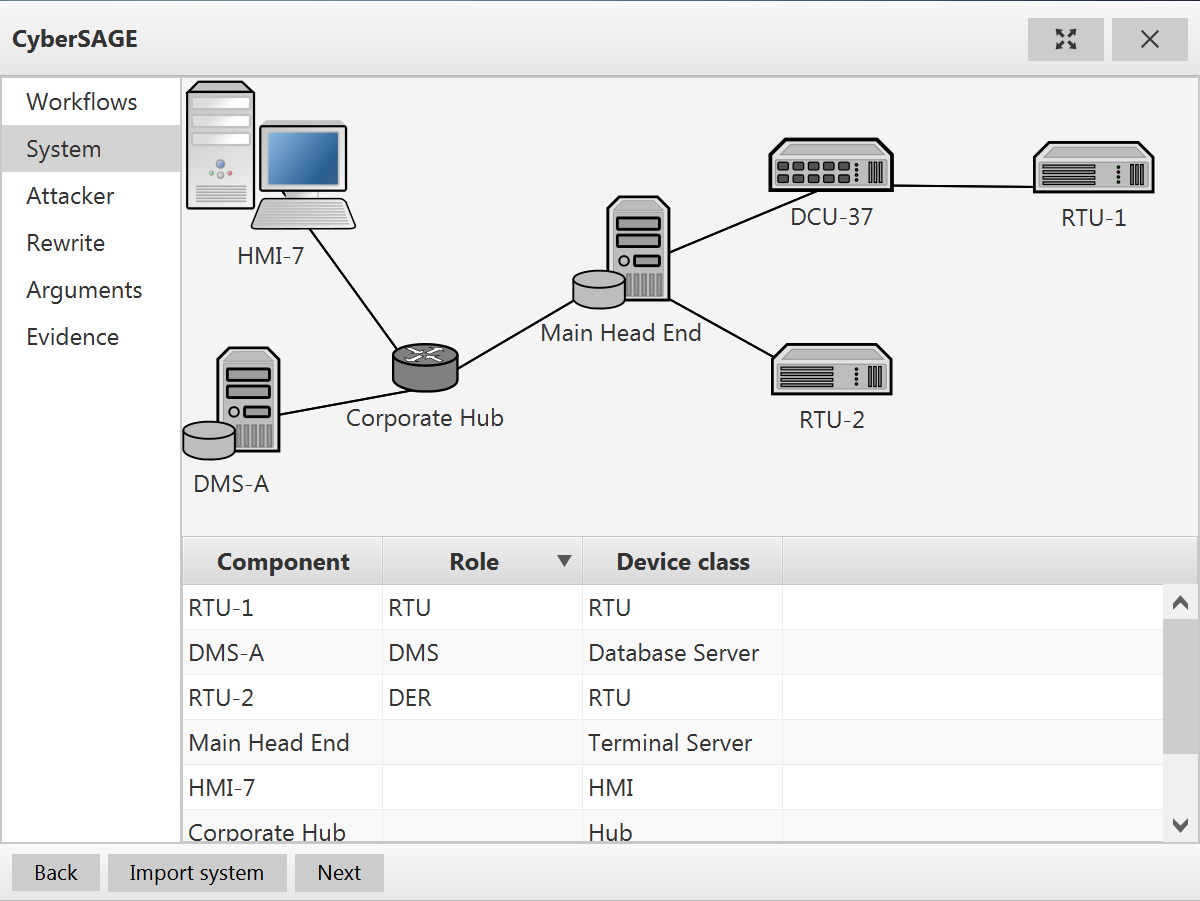}
\caption{}
\label{fig:cybersage1}
\end{subfigure}
\begin{subfigure}{0.6\textwidth}
\centering
\includegraphics[width=\linewidth]{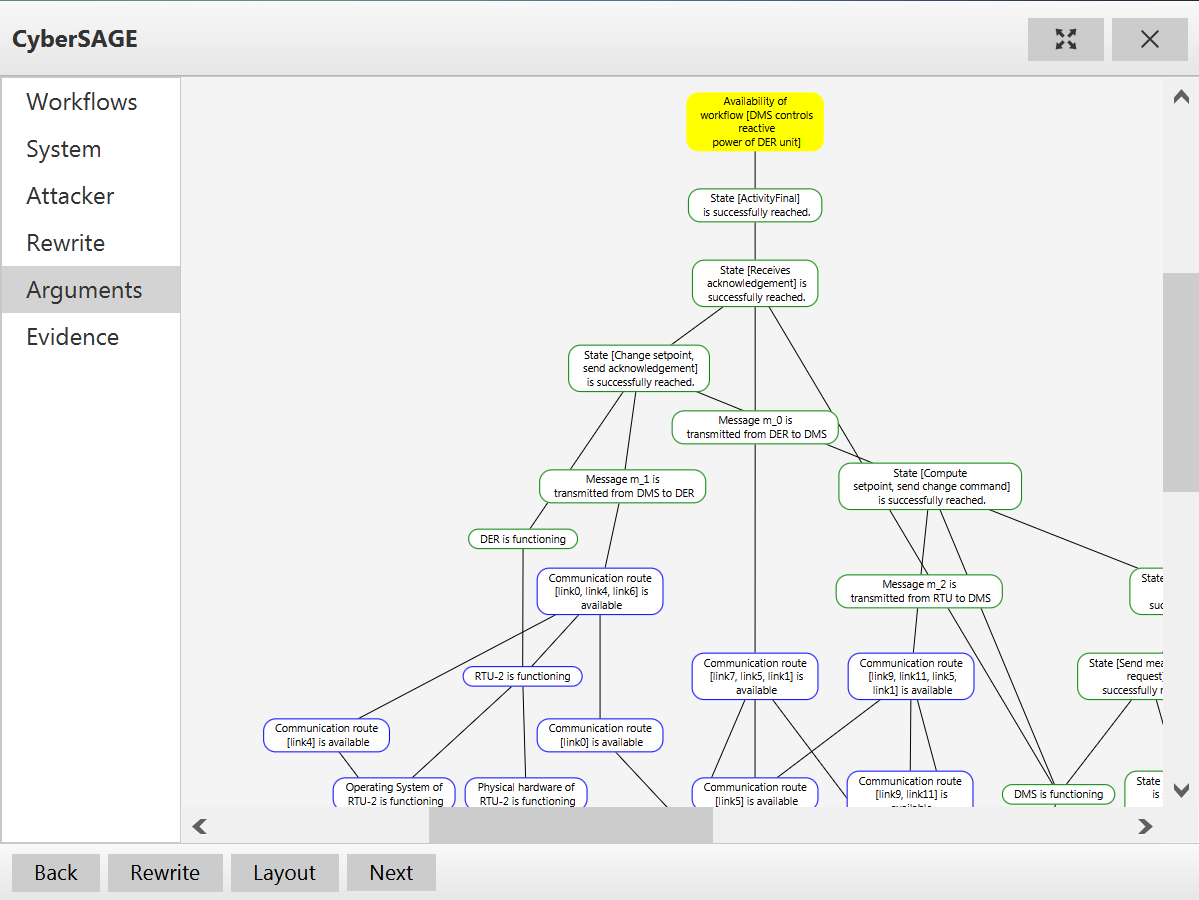}
\caption{}
\label{fig:cybersage2}
\end{subfigure}
\caption{CyberSAGE prototype: (a) System input configuration (topology, device information, mapping of actors to devices). (b) Part of the GSA graph, automatically generated using the extension templates introduced in this paper.}
\label{fig:screenshot}
\end{figure*} 

We implemented the proposed template-based graph generation using our prototype assessment tool, which we call CyberSAGE (Cyber Security Argument Graph Evaluation)~\cite{cybersage}. Based on the inputs described in Section~\ref{sec:usecase}, CyberSAGE is able to automatically generate the argument graph. Screen-shots of the system input and part of the GSA-graph are
shown in Fig.~\ref{fig:cybersage1} and Fig.~\ref{fig:cybersage2}, respectively. 

The workflow and system inputs are given to CyberSAGE in XML
format. Currently, our tool does not provide functionality to edit
these inputs; it depends on other applications, such as XML/text
editors, to obtain and modify them. For example, CyberSAGE imports the
system topology input directly from save files of the CSET
tool\footnote{The Cyber Security Evaluation Tool: \url{ics-cert.us-cert.gov/satool.html}}.  Once CyberSAGE has
imported those files, it builds its internal data structures and
visualizes their contents in a user-friendly manner (see
Fig.~\ref{fig:cybersage1}).

The remaining inputs, namely the attacker models and
extension templates, are currently static and provided by CyberSAGE.
With the extension templates implemented, CyberSAGE automatically
builds the argument graph by following Algorithm~\ref{alg:graphA}. It
takes the tool less than 1 second
to produce the final argument graph, which contains around 50 vertices. The tool
also provides some degree of customization, such as allowing the user to
enable or disable a particular subset of extension templates when
building the graph.

\section{Related Work}
\label{sec:related}
We see parallels between our work on security argument graph generation and work from the safety and reliability communities, as well as related efforts within the security domain. In this section, we discuss related efforts in graph-based modeling and highlight unique features of our framework. 

\Paragraph{Safety case generation}
A safety case uses certain argument strategies to organize a body of evidence so as to provide a compelling case for supporting certain safety  claims (goals)~\cite{kelly98}. Safety cases are typically constructed manually. Recent efforts (e.g., ~\cite{denney12,denney13}) have begun to introduce formal semantics to help automate the safety case generation process.
Compared with those recent efforts, our proposed approach focuses on argument patterns that incorporate various security-related evidence, including security goals and attacker models. We also formalize the template in a local way, which simplifies the definition and instantiation of the template while still allowing progressive generation of the argument graph.

\Paragraph{Fault tree generation}
Fault tree analysis is a classic deductive method used to determine what combinations of basic component failures can lead to a system-level fault event~\cite{vesely81fault}. While fault trees are usually constructed manually in practice, there has been a steady stream of efforts to automate the fault tree generation process. For example, 
Pai et al.~\cite{pai02} propose a method to transform a UML system model to dynamic fault trees. Joshi et al.~\cite{joshi07} propose a method to automatically generate a dynamic fault tree from an Architectural Analysis
and Design Language (AADL) model. Recently, Xiang et al.~\cite{xiang11} propose an automatic synthesis method to generate a static fault tree from a system model specified with SysML.

Compared with that line of work, our proposed approach not only considers various security-specific information, but also intends to cover a broader scope of security-related claims and heterogeneous pieces of evidence.


\Paragraph{Attack trees and other security assessment techniques}
Attack trees and their variations (e.g., attack graphs~\cite{sheyner02,ou06}, ADVISE~\cite{lemay11},  and attack-defense trees~\cite{kordy10adt}) have been shown to be useful for security assessment. Inspired by the fault tree formalism, an attack tree graphically represents how a potential threat can be realized through various possible combinations of attacks. Attack trees are usually constructed manually. For the specific domain of network security, multiple efforts (e.g.,~\cite{sheyner02,ou06}) have tried to automate the generation of attack graphs, which are meant to model how an attacker can use staged attacks to compromise certain assets in a network. ADVISE~\cite{lemay11} automates the search of attack strategies. Those efforts differ from ours in that they do not provide a framework that can automatically integrate heterogeneous pieces of information (e.g., relating to security goals, workflows, system information, and the attacker) to produce a holistic security argument. 



\section{Conclusion}
\label{sec:conclusion}

In this work, we consider graph-based security assessment for complex systems. 
Such assessments can be used to identify
potential attacks and to compare the security properties of different
designs. Because of the size of real-world systems, the assessment process should ideally be tool-supported and automated to a large
extent. 
In~\cite{Chen2013} we proposed a holistic security assessment framework that combines
a collection of heterogeneous input information into a security argument graph. Building on that work, in this paper we identify a collection of argument patterns that emerge from relationships between input classes. We formalize these patterns as extension templates, and provide a method for automated graph generation using these templates. We describe how we implemented the proposed method in our security assessment tool, CyberSAGE, and demonstrate the automated generation process for an example use case from the power sector. 

\section*{Acknowledgments}
This work is supported by Singapore's Agency for Science, Technology, and Research (A*STAR) under the Human Sixth Sense Programme (HSSP). We thank Jenny Applequist for her help with editing.

\balance

\bibliographystyle{abbrv}
\bibliography{references}

\begin{thebibliography}{10}

\bibitem{amin13cyber1}
S.~Amin, X.~Litrico, S.~Sastry, and A.~M. Bayen.
\newblock Cyber security of water {SCADA} systems - part {I}: Analysis and
  experimentation of stealthy deception attacks.
\newblock {\em IEEE Trans. Contr. Sys. Techn.}, 21(5):1963--1970, 2013.

\bibitem{anderson10}
R.~Anderson and S.~Fuloria.
\newblock Who controls the off switch?
\newblock In {\em Proc. of the IEEE Conference on Smart Grid Communications
  (SmartGridComm)}, 2010.

\bibitem{etsi-architecture}
{CEN-CENELEC-ETSI Smart Grid Coordination Group}.
\newblock Smart grid reference architecture.
\newblock
  \url{http://ec.europa.eu/energy/gas_electricity/smartgrids/doc/xpert_group1_reference_architecture.pdf},
  November 2012.

\bibitem{Chen2013}
B.~Chen, Z.~Kalbarczyk, D.~M. Nicol, W.~H. Sanders, R.~Tan, W.~G. Temple, N.~O.
  Tippenhauer, A.~H. Vu, and D.~K. Yau.
\newblock Go with the flow: Toward workflow-oriented security assessment.
\newblock In {\em Proc. of the New Security Paradigms Workshop (NSPW)}, 2013.

\bibitem{denney13}
E.~Denney and G.~Pai.
\newblock A formal basis for safety case patterns.
\newblock In {\em Proc. of the Conference on Computer Safety, Reliability and
  Security (SAFECOMP)}, 2013.

\bibitem{denney12}
E.~Denney, G.~Pai, and J.~Pohl.
\newblock Heterogeneous aviation safety cases: Integrating the formal and the
  non-formal.
\newblock In {\em Proc. of the Conference on Engineering of Complex Computer
  Systems (ICECCS)}, 2012.

\bibitem{joshi07}
A.~Joshi, S.~Vestal, and P.~Binns.
\newblock Automatic generation of static fault trees from {AADL} models.
\newblock In {\em Proc. of DSN Workshop on Architecting Dependable Systems},
  2007.

\bibitem{kelly98}
T.~Kelly.
\newblock {\em Arguing Safety: A systematic Approach to Managing Safety Cases}.
\newblock PhD thesis, University of York, UK, 1998.

\bibitem{kordy10adt}
B.~Kordy, S.~Mauw, S.~Radomirovi\'{c}, and P.~Schweitzer.
\newblock Foundations of attack-defense trees.
\newblock In {\em Proc. of the conference on Formal Aspects of Security and
  Trust (FAST)}, pages 80--95, 2011.

\bibitem{kordy2013survey}
B.~Kordy, L.~Pietre-Cambacedes, and P.~Schweitzer.
\newblock {DAG}-based attack and defense modeling: Don't miss the forest for
  the attack trees.
\newblock {\em CoRR}, abs/1303.7397, 2013.

\bibitem{lemay11}
E.~LeMay, M.~Ford, K.~Keefe, W.~H. Sanders, and C.~Muehrke.
\newblock {Model-based security metrics using ADversary VIew Security
  Evaluation (ADVISE)}.
\newblock In {\em Proc. of the Conference on Quantitative Evaluation of SysTems
  (QEST)}, 2011.

\bibitem{nicol04}
D.~M. Nicol, W.~H. Sanders, and K.~S. Trivedi.
\newblock Model-based evaluation: from dependability to security.
\newblock {\em IEEE Transactions on Dependable and Secure Computing},
  1(1):48--65, 2004.

\bibitem{ou06}
X.~Ou, W.~Boyer, and M.~McQueen.
\newblock A scalable approach to attack graph generation.
\newblock In {\em Proc. of the ACM Conference on Computer and Communications
  Security (CCS)}. ACM, 2006.

\bibitem{pai02}
G.~J. Pai and J.~B. Dugan.
\newblock Automatic synthesis of dynamic fault trees from {UML} system models.
\newblock In {\em Proc. of the Symposium on Software Reliability Engineering
  (ISSRE)}, 2002.

\bibitem{schneier99}
B.~Schneier.
\newblock Attack trees: Modeling security threats.
\newblock {\em Dr. Dobb's Journal}, 1999.

\bibitem{sheyner02}
O.~Sheyner, J.~Haines, S.~Jha, R.~Lippmann, and J.~Wing.
\newblock Automated generation and analysis of attack graphs.
\newblock In {\em Proc. of the IEEE Symposium on Security and Privacy}, 2002.

\bibitem{temple13delay}
W.~G. Temple, B.~Chen, and N.~O. Tippenhauer.
\newblock Delay makes a difference: Smart grid resilience under remote meter
  disconnect attack.
\newblock In {\em Proc. of the IEEE Conference on Smart Grid Communications
  (SmartGridComm)}, 2013.

\bibitem{verendel09}
V.~Verendel.
\newblock Quantified security is a weak hypothesis.
\newblock In {\em Proc. of the New Security Paradigms Workshop (NSPW)}, 2009.

\bibitem{vesely81fault}
W.~E. Vesely, F.~F. Goldberg, N.~H. Roberts, and D.~F. Haasl.
\newblock {\em {Fault Tree Handbook}}.
\newblock U.S. Nuclear Reg. Comm., 1981.

\bibitem{cybersage}
A.~H. Vu, N.~O. Tippenhauer, B.~Chen, D.~M. Nicol, and Z.~Kalbarczyk.
\newblock Cybersage: A tool for automatic security assessment of cyber-physical
  systems.
\newblock In {\em Proc. of the Conference on Quantitative Evaluation of SysTems
  (QEST)}, 2014.

\bibitem{xiang11}
J.~Xiang, K.~Yanoo, Y.~Maeno, and K.~Tadano.
\newblock Automatic synthesis of static fault trees from system models.
\newblock In {\em Proc. of the Conference on Secure Software Integration and
  Reliability Improvement (SSIRI)}, 2011.

\end{thebibliography}

\end{document}